\documentclass[apj]{emulateapj}

\usepackage{amssymb,amsmath}
\usepackage{natbib}
\shorttitle{Color dependence of clustering of massive galaxies at 0.5$\le z \le$2.5}
\shortauthors{Lin et al.}

\begin{document}
	
\title{Color dependence of clustering of massive galaxies at 0.5$\le z \le$2.5: similar spatial distributions between green valley galaxies and AGNs}

\author{Xiaozhi Lin\altaffilmark{1,2,3,4}, Guanwen Fang\altaffilmark{1,7}, Zhen-Yi Cai\altaffilmark{3,4}, Tao Wang\altaffilmark{5}, Lulu Fan\altaffilmark{6}, Xu Kong\altaffilmark{3,4}}

\altaffiltext{1}{Institute for Astronomy and History of Science and Technology, Dali University, Dali 671003, China; wen@mail.ustc.edu.cn}
\altaffiltext{2}{Shandong Provincial Key Laboratory of Optical Astronomy and Solar-Terrestrial Environment, Shandong University, Weihai, 264209}
\altaffiltext{3}{CAS Key Laboratory for Research in Galaxies and Cosmology, Department of Astronomy, University of Science and Technology of China, Hefei 230026, China; xkong@ustc.edu.cn}
\altaffiltext{4}{School of Astronomy and Space Science, University of Science and Technology of China, Hefei 230026, China}
\altaffiltext{5}{Laboratoire AIM-Paris-Saclay, CEA/DSM/Irfu, CNRS, Université Paris Diderot, Saclay, pt courrier 131, 91191 Gifsur-Yvette, France}
\altaffiltext{6}{Shandong Provincial Key Lab of Optical Astronomy and Solar-Terrestrial Environment, Institute of Space Sciences, Shandong University, Weihai, 264209, China}
\altaffiltext{7}{Guanwen Fang and Xiaozhi Lin contributed equally to this work.}

\begin{abstract}
We present a measurement of the spatial clustering of rest-frame UV-selected massive galaxies at $0.5\le z \le 2.5$ in the COSMOS/UltraVISTA field. Considering four separate redshift bins with $\Delta z=0.5$, we construct three galaxy populations, i.e., red sequence (RS), blue cloud (BC), and green valley (GV) galaxies, according to their rest-frame extinction-corrected UV colors. The correlation lengths of these populations are confirmed to be dependent on their rest-frame UV color and redshift: UV redder galaxies are found to be more clustered. In all redshift bins, the GV galaxies generally have medium clustering amplitudes and are hosted within dark matter halos whose masses are more or less between those of RS and BC galaxies; and the clustering amplitude of GV galaxies is close to that of AGNs in the same redshift bin, suggesting that AGN activity may be responsible for transforming galaxy colors. After carefully examining their stellar masses, we find that the clustering amplitudes of galaxy samples with different colors are all similar once they have a similar median stellar mass and that the median stellar mass alone may be a good predictor of galaxy clustering.
\end{abstract}


\section{Introduction}

Many studies have been statistically exploring and comparing the properties of early-type red quiescent galaxies (QGs) and late-type blue star-forming galaxies (SFGs) at different redshifts \citep[e.g.,][]{Whitaker_2011,Patel_2013}. The fact that QGs are generally red while SFGs are blue is coined as the so-called color bimodality. This color bimodality has been observed in the local universe \citep{Strateva_2001,Blanton_2003}, out to $z\sim1$ by the COMBO-17 survey \citep{Bell_2004,Borch_2006}, up to $z\sim1.5$ by the VIMOS-VLT deep survey \citep{Franzetti_2007} and by the UKIDSS Ultra Deep Survey \citep{Cirasuolo_2007}, and plausibly up to $z\sim2$ in the Hubble Deep Field \citep{Giallongo_2005} and by the GMASS project \citep{Cassata_2008}. Recently, owing to the innovation of the large and deep field surveys, the color bimodality is revealed up to $z\sim3$ \citep{Whitaker_2011}. Moreover, the color bimodality has been reported using various colors from ultraviolet to infrared (e.g., $g-r$, \citealt{Blanton_2003}; $u-r$, \citealt{Baldry_2004}; ${U-B}$, \citealt{Cassata_2008}; ${\rm NUV}-r$, \citealt{Wyder_2007}; $[3.4~\mu m]-[12~\mu m]$, \citealt{Lee_2015}).

Although some dust hidden SFGs may contaminate QGs \citep{Papovich_2012,Williams_2009}, the color bimodality has usually been used to separate SFGs from QGs. The latter with older stellar populations are mostly found in denser environments, more concentrated in luminosity, and exhibit more symmetric in morphology \citep{Whitaker_2013,Pan_2013}, thus the physical properties of these two populations appear differently and distinct formation mechanisms may be suggested by this color bimodality \citep{Scarlata_2007,Delucia_2007}.

Furthermore, galaxies at high redshifts are found to be dominated by SFGs while QGs are more common in the local universe \citep{Wuyts_2011,Elbaz_2011}. The sizes of QGs at high redshifts are significantly smaller than those of QGs in the local universe \citep{Newman_2012,Bruce_2012,Szomoru_2013}. Interestingly, \citet{Barro_2013} discovered a population of massive SFGs at redshifts $z=1.5-3$ with very compact structures being distinct from normal extended SFGs. Since the number densities, masses, sizes, and star formation rates (SFRs) of these massive compact star-forming galaxies (cSFGs) are all close to those of massive compact quiescent galaxies (cQGs) at redshifts $z=1.5-3$, they proposed two evolutionary tracks of QG formation: an early path at $z \gtrsim 2$ that, as a result of gas-rich processes such as mergers or disk-instabilities, massive cSFGs formed at $z = 2-3$ may be quenched by gas exhaustion, stellar and/or active galactic nuclei (AGNs) feedbacks and quickly evolve into massive cQGs at $z \gtrsim 1.5$, thereafter, these cQGs formed at high redshifts may continuously enlarge their sizes toward the present; and a late path at $z \lesssim 2$ that, as a result of secular processes, halo quenching, or gas-poor mergers, extended QGs are directly formed from extended SFGs.

However, the formation and size evolution of QGs and their relationship with SFGs have not yet been clearly understood \citep{Barro_2013,Williams_2014}. To further explore their relationship and any possible transitional population, using the color-mass relation \citep{Bell_2004,Borch_2006}, the whole galaxies are usually separated into different sequences, i.e., a red sequence (RS) for QGs, a blue cloud (BC) for SFGs, and a transition zone between them for the so-called green valley (GV) galaxies \citep[e.g.,][]{Coil_2008,Mendez_2011,Wang_2017}. Many properties of GV galaxies also lie between those of RS and BC galaxies, including their clustering properties \citep{Coil_2008,Heinis_2009,Loh_2010}, morphologies \citep{Mendez_2011}, luminosity functions \citep{Goncalves_2012}, dusts and SFRs \citep{Salim_2014}, gas properties \citep{Schawinski_2014}, environments \citep{Lee_2015}, and AGN hosting fractions \citep{Wang_2017}. In this work, we mainly focus on the clustering properties of these three populations in order to explore the role of GV galaxies in massive galaxy evolution and the relationship to AGN activities up to very high redshift.

The clustering analysis provides us with a tool to associate the properties of galaxies with large-scale structures of the universe and could shed light on their evolution. Historically, the correlation between the color bimodality and clustering amplitude was first established by \citet{Davis_1976}, who found that elliptical galaxies are more tightly clustered than spiral galaxies. The conclusion that RS galaxies have a higher clustering amplitude than BC galaxies was confirmed by a number of subsequent studies in the local universe \citep{Loveday_1995,Willmer_1998} and up to $z \sim 1.5$, thanks to the 2dF Galaxy Redshift Survey \citep{Norberg_2002,Madgwick_2003}, the early Deep Evolutionary Exploratory Probe 2 (DEEP2) Galaxy Redshift Survey \citep{Coil_2004}, the Sloan Digital Sky Survey \citep[SDSS;][]{Zehavi_2011}, the first epoch VIMOS-VLT Deep Survey \citep{Meneux_2006}, and many others \citep[e.g.,][]{Shepherd_2001,Firth_2002}. Recently, \citet{Coil_2017} confirmed the same result out to $z \sim 1$ using a sample of over 100,000 spectroscopic redshifts from the PRIsm Multi-object Survey (PRIMUS) and DEEP2 galaxy redshift surveys.

The clustering study of GV galaxies firstly appeared in \citet{Coil_2008} who used the DEEP2 data to study the clusterings of RS, BC, and GV galaxies at $z\sim1$, and found that the clustering of GV galaxies lies between those of RS and BC populations. Similar conclusions were obtained by following studies in the local universe and up to $z \sim 1$ \citep{Loh_2010,Zehavi_2011,Bray_2015}. At $0.2 < z < 1.0$, \citet{Dolley_2014} found that many galaxies with the highest SFRs are likely the GV galaxies and have stronger clustering than the typical BC galaxies.

All in all, this color dependence of clustering has been established for RS and BC galaxies, while less obvious for GV galaxies below $z \sim 1$ \citep[e.g.,][]{Coil_2008,Loh_2010,Zehavi_2011,Bray_2015}, partly due to the different operational definitions of the GV galaxies used in different studies. Here we would extend these clustering analyses for RS, GV, and BC galaxies up to $z \sim 2.5$, adopting a newly refined selection method by \citet{Wang_2017}.

Furthermore, what a role GV galaxies play in massive galaxy evolution at higher redshift is still an interesting question under debate. The clustering of galaxies with different colors is also found to be related to that of AGNs they host. For example, by studying the clustering of galaxy and AGN samples at $0.25<z<0.8$ in the AGES field, \citet{Hickox_2009} find that the X-ray-selected AGNs are preferentially found in GV galaxies. Since AGN feedback may be responsible for the evolution of BC to RS galaxies, GV galaxies as a transition population may be related to the AGN activity.

In this paper, we analyze the clustering properties of massive RS, GV, and BC galaxies at $0.5 \le z \le 2.5$. The three galaxy populations are firstly constructed in Section~\ref{sect:data} and then we calculate their correlation lengths and the corresponding halo masses in Section~\ref{sect:clustering}. By comparing with the clustering properties of cSFGs and massive QGs, we explore and discuss a possible evolution sequence among these galaxies at high redshift in Section~\ref{sect:RaD}, followed by a short summary in Section~\ref{sect:summary}.

All our data are from the COSMOS/UltraVISTA survey, whose depth and large size enable us to make an accurate and reliable clustering analysis. Throughout this paper, we adopt a flat cosmology with \(\Omega_{m} = 0.3,\) \(\Omega_{\Lambda} = 0.7,\) and \(H_{0} = 70\hspace{0.5mm}{\rm km~s^{-1}~Mpc^{-1}}\). We assume a normalization of \(\sigma_{8} = 0.84\) for the matter power spectrum. All quoted uncertainties are 1$\sigma$ (68\% confidence).

\section{Data}\label{sect:data}

The COSMOS/UltraVISTA survey has a wealth of imaging data and covers various wavelengths from ultraviolet to mid-infrared. In our study, we use the catalog covering $\sim 1.62$ deg$^2$ of the COSMOS/UltraVISTA field \citep[containing 262,615 sources down to the 3$\sigma$ limit of $K_{\rm s} < 24.35$;][]{Muzzin_2013a}. The catalog is constructed from the UltraVISTA $K_s$ band imaging \citep[][]{McCracken_2012} and includes multi-wavelength photometries (i.e., 30 bands from $\sim 0.15~\mu$m to $\sim 24~\mu$m), photometric redshifts, rest-frame colors, and stellar population parameters, estimated by the spectral energy distribution (SED) fitting method \citep[cf.][for details]{Muzzin_2013a}.

\subsection{Redshift, Stellar Mass, and $A_{V}$}

We directly adopt the photometric redshifts and rest-frame colors estimated by \citet{Muzzin_2013a} for our analysis. The relevant data and the deduction of photometric redshifts are briefly presented in the following together with the quality of photometric redshifts \citep[see][for more details]{Muzzin_2013a}.

The observed-frame photometric catalog from \citet{Muzzin_2013a} contains two bands from GALEX (FUV and NUV), one band from CFHT/MegaCam ($u^{*}$), six broad bands taken with Subaru/SuprimeCam ($g^{+}r^{+}i^{+}z^{+}{\rm B}_{j}{\rm V}_{j}$), 12 optical medium bands (IA427-IA827) from Subaru/SuprimeCam, four near-infrared broad bands ({\it YJHK$_{s}$}) as well as the 3.6 $\mu$m, 4.5 $\mu$m, 5.8 $\mu$m, 8.0 $\mu$m, and 24 $\mu$m channels from {\it Spitzer}'s IRAC+MIPS cameras. The observed-frame wavelength coverage of this catalog is from $\sim 0.15~\mu $m to $\sim 24~\mu$m, and then for our considered maximal redshift of $\sim 2.5$ the rest-frame wavelength coverage is $\sim 0.04~\mu$m to $\sim 6.8~\mu$m. Therefore, the UltraVISTA data have a good coverage in the redshift desert between $z \sim 1.4-2.5$.

The photometric redshifts and rest-frame colors in the COSMOS/UltraVISTA catalog are computed using EAZY \citep{EAZY}. EAZY determines the $z_{\rm phot}$ for galaxies by fitting their SEDs to a linear combination of several galactic templates. Besides the initial seven templates in EAZY (i.e., six from the PEGASE models of \citealt*{FR_1999} and one red template from the model of \citealt*{M05}), in order to improve the accuracy of photometric redshift, \citet{Muzzin_2013a} further consider two more templates, that is, a one-gigayear-old single-burst \citet{BC03} model for strong post-starburst-like features at $z>1$, and a slightly dust-reddened young population for the UV bright galaxies at $1.5<z<3.5$. We note that no AGN template has been included when estimating the photometric redshift, because we will exclude all sources that are considered to be bright stars or quasars in the \citet{Muzzin_2013a} catalog.

The resultant $z_{\rm phot}$ at $z < 1.5$ consists well with $z_{\rm spec}$ from zCOSMOS \citep{Lilly_2007,Lilly_2009} and other spectroscopic surveys \citep{Onodera_2012,Bezanson_2013,Sande_2013}, with low $3\sigma$-outlier fraction of $1.56\%$ and low rms scatter of $\delta z/(1+z)=0.013$. At $z > 1.5$, the $z_{\rm phot}$ also agrees well with $z_{\rm phot}$ determined using the NOAO Extremely Wide-Field Infrared Imager (NEWFIRM) Medium-Band Survey \citep[NMBS,][]{Whitaker_2011}, with the fraction of the UltraVISTA sources considered to be $5\sigma$ outliers of the NMBS survey to be $2.0\%$ and the rms dispersion of UltraVISTA photometric redshifts to be $\delta z/(1+z)=0.026$.

There is indeed a small population of extreme blue galaxies that are confusedly fitted. Their $P(z)$ distributions are frequently bi-modal with a peak at $z<0.5$ as well as a peak at $z>1.5$. Their SEDs are also typically very blue with a single breaking feature in the bluest optical bands. This breaking feature can be interpreted as the Balmer break at $z<0.5$ or the Lyman break at $z>1.5$. Based on careful examination of their SEDs and spectra if available, they are expected to be high-redshift galaxies. Since the NMBS survey only covers a fraction of the UltraVISTA field, \citet{Muzzin_2013a} adopt a somewhat ad hoc correction based on the $UVJ$ diagram for this population. However, thanks to only a tiny fraction ($<1\%$) of the total sample is affected by this correction, and more importantly, almost all of them lie below the mass-completeness limit, our following analyses based on the mass-complete sample are not expected to be affected by the correction.

Although a $\chi^2$ value assessing the SED fitting is provided by \citet{Muzzin_2013a} for each galaxy, we do not apply any cut on this value since nearly $\sim 90\%$ of the sample has a reduced $\chi^2 < 2$.

For the purpose of our study, removing the star/quasar contamination using the flag in the catalog and further limiting galaxies at $0.5\le z_{\rm phot}\le 2.5$, we obtain 126,222 sources, which are further divided into four redshift bins, that is, $0.5\le z<1.0$, $1.0\le z <1.5$, $1.5\le z <2.0$, and $2.0 \le z \le2.5$, in order to study the evolution of galactic clustering properties.

Instead, the stellar mass, $M_{\ast}$, and visual attenuation, $A_{V}$, of galaxies are computed using FAST code by fitting the observed SEDs to \citet{M05} models assuming solar metallicity, a Chabrier initial mass function \citep[][]{Chabrier_2003}, and a \citet{Calzetti_2000} dust extinction law. Exponentially declining star formation histories are assumed with the e-folding star formation timescales ranging from $10^{7}\hspace{0.5mm}{\rm yr}$ to $10^{10}\hspace{0.5mm}{\rm yr}$.

\begin{figure}
\centering
\includegraphics[width=\columnwidth, angle=0]{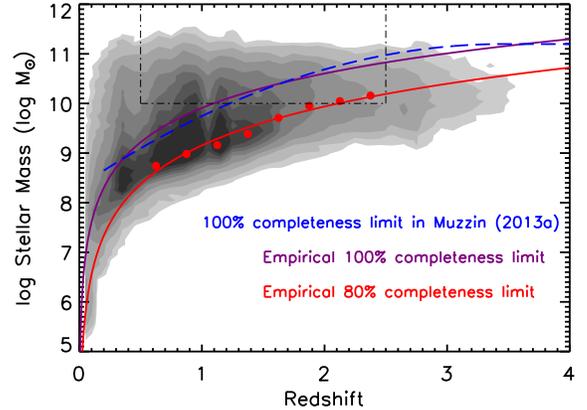}
\caption{The stellar masses of all COSMOS/UltraVISTA galaxies as a function of redshift. We empirically derive 100\% (blue solid line) and 80\% (red solid line and filled circles) mass-completeness limits by scaling down galaxies to the depth of ${K}_s=23.4$. Also shown is the 100\% mass-completeness limit in \citet[blue dashed line,][]{Muzzin_2013b}. We restrict our sample to $0.5\le z_{\rm phot}\le 2.5$ and $\log(M_{\ast}/M_{\odot}) \ge 10$ (i.e., within the dot-dashed rectangle).}
\label{fig:mass_completeness}
\end{figure}

\subsection{Sample Construction}\label{sect:sample_construction}

The following analysis only consists of the COSMOS/UltraVISTA galaxies with $\log(M_{\ast}/M_{\odot}) \ge 10$ (totally 41,218 sources) such that the mass-completeness limit in the concerned redshift range can reach about $80\%$. These attractive galaxies are enclosed by the dot-dashed rectangle illustrated in Figure~\ref{fig:mass_completeness} for the stellar mass of all COSMOS/UltraVISTA galaxies versus redshift.

The statement on the mass-completeness limit is obtained adopting an empirical method to estimate the redshift-dependent mass-completeness limit following \citet{Quadri_2012}. Briefly speaking, we select a brighter sub-sample at $0.5\le z\le 2.5$ that is a factor of 1.0 - 2.0 above the 90\% completeness limit \citep[$K_{s}=23.4$,][]{Muzzin_2013a}. In each smaller redshift bin of $\Delta z=0.25$, we scale their fluxes and stellar masses down to the adopted limit of $K_{s} = 23.4$ by the same factors, and then define the 100\% and 80\% mass-completeness limits by encompassing the 100\% and 80\% of the brighter sub-sample, respectively (cf. the red filled circles in Figure~\ref{fig:mass_completeness} for the 80\% mass-completeness limit as a function of redshift). Thereafter, these discrete mass-completeness limits are then fit by parametrization enveloping lines, $M_{\rm lim}=a+b\ln(z)$. As a result, the 100\% and 80\% mass-completeness limits can be described as $9.91+0.96\ln(z)$ and $9.15+1.13\ln(z)$, respectively. The 100\% mass-completeness limit agrees well with that of \citet{Muzzin_2013b}. Although the mass-completeness limit reaches 100\% as far as $\log(M_{\ast}/M_{\odot}) \ge 11$, the resultant sample size would be too small to get a meaningful clustering analysis. Therefore, we consider galaxies with $\log(M_{\ast}/M_{\odot}) \ge 10$ such that the mass-completeness limit is still above 80\% and redshifts can be up to $z\sim2.5$.

To check the effect of the incompleteness on the clustering results, we re-sample our galaxies with stellar mass $\log(M_{\ast}/M_{\odot}) \ge 10.5$, which corresponds to the 98\% completeness limit. Similar clustering analyses in the aforementioned four redshift bins are performed on these samples for comparison. If we increase the stellar mass threshold up to $10^{11}~M_{\odot}$, the total source numbers of RS, BC, and GV galaxies (see following for the separation criteria) will decrease to 1692, 219, and 284, respectively. In the latter case, we only calculate the clustering amplitudes of RS galaxies at the three redshift bins of $0.5\le z<1.0$, $1.0\le z<1.5$, and $1.5\le z<2.0$, because the source numbers of RS galaxies in the highest redshift bin and the other two galaxy samples with $\log(M_*/M_{\odot}) \ge 11$ are all less than 100, then the corresponding angular correlation functions would be unreliable. The effects of stellar mass threshold on the clustering will be further discussed in Section~\ref{sect:RaD}.

\begin{figure}
\centering
\includegraphics[width=0.8\columnwidth, angle=90]{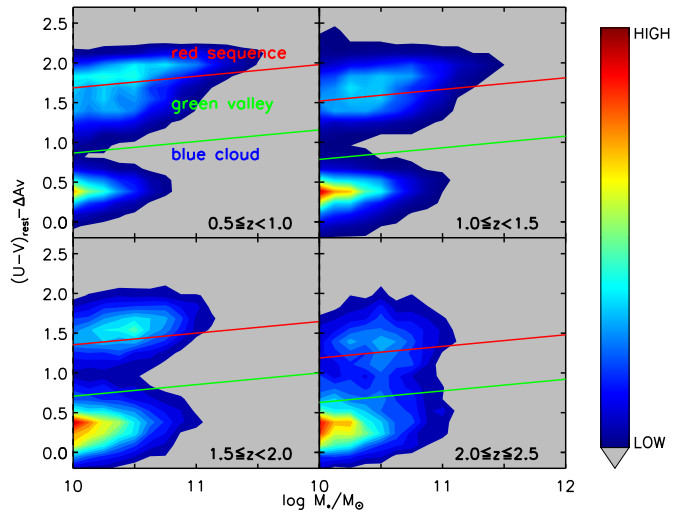}
\caption{The distributions of the concerned COSMOS/UltraVISTA galaxies with $0.5\le z\le 2.5$ and $\log(M_{\ast}/M_{\odot}) \ge 10$ in the color-mass diagram with the de-reddened rest-frame UV color. Denser regions are redder. At the center of each redshift bin, the red and green lines roughly depict the separations of RS, GV, and BC galaxies, under the adopted separation criteria (see Equation~\ref{ref:equ_sep_cri}).}
\label{fig:CMR}
\end{figure}

Since we are concerning of the clusterings of RS, GV, and BC galaxy populations, following \citet{Wang_2017}, we apply their definitions in the rest-frame extinction-corrected color-mass diagram to construct three sub-samples for the RS, GV, and BC galaxy populations as shown in Figure~\ref{fig:CMR}. \citet{Wang_2017} upgrade the old separation criteria proposed by \citet{Bell_2004} and \citet{Borch_2006} such that GV galaxies can be more successfully separated from RS and BC galaxies and that the dusty BC galaxies can also be better separated from RS galaxies. Although the old criteria have been shown to be valid up to $z \sim 3$ in separating the BC galaxies from the RS ones \citep{Xue_2010}, it is derived using the observed colors without extinction-correction and therefore the RS galaxies may be contaminated by the real BC galaxies but with large dust extinction. Instead, as also shown by \citet{Brammer_2009}, using the extinction-corrected $U-V$ color can better separate RS galaxies due to dust attenuation from those due to old stellar populations and therefore galaxies separated in this way would have more clear distinct star formation properties. Moreover, \citet{Wang_2017} check out the RS, GV, and BC galaxy populations divided by the new separation criteria, and find that they are more self-consistent from local to high redshifts.

The adopted separation criteria are described as
\begin{equation}\label{ref:equ_sep_cri}
\begin{split}
{\rm RS:(UV)_{rest}} \ge 0.126 \log(M_{\ast}/M_\odot)+0.58-0.286z\\
{\rm GV:(UV)_{rest}} < 0.126 \log(M_{\ast}/M_\odot)+0.58-0.286z\\
\&\hspace{1mm}({\rm UV)_{rest}} \ge 0.126 \log(M_{\ast}/M_\odot)-0.24-0.136z\\
{\rm BC:(UV)_{rest}} < 0.126 \log(M_{\ast}/M_\odot)-0.24-0.136z,
\end{split}
\end{equation}
where ${\rm(UV)_{rest}}$ is the rest-frame de-reddened UV color that satisfies ${\rm (UV)_{rest}} \equiv (U-V)_{\rm rest}- \Delta \times A_{V}$, $(U-V)_{\rm rest}$ is the rest-frame observed UV color, and $\Delta = 0.47$ is the correction factor computed for the \citet{Calzetti_2000} extinction law. By applying this correction factor the dust hidden SFGs can be better separated from red and dead galaxies as demonstrated by \citet{Brammer_2009}.
Every galaxy is adjudged under the above criteria and assigned to one of the three galaxy populations, i.e., RS, GV, or BC.
The source number and mean redshift of each galaxy population within each redshift bin are tabulated in Table~\ref{tab:galaxy_bin}, respectively.

\section{Clustering analysis}\label{sect:clustering}

In this section, we analyze the angular and spatial clusterings for our RS, GV, and BC samples. Using these methods, we try to place the GV sample in the galaxy evolution sequence and link it to the other two galaxy populations. In Section~\ref{sect:acf}, we study the two-point angular correlation functions of each sample in different redshift intervals, while the corresponding properties of the spatial clustering are estimated in Section~\ref{sect:sc}. Then we assess the host dark matter (DM) halo masses and the bias factors for our samples in Section~\ref{sect:bias}. Finally, in order to compare with the observed spatial clustering of galaxies, Section~\ref{sect:clustering_model} presents a recipe for approximately modeling the evolution of spatial clustering of galaxies hosted within DM halos with a constant or growing mass.

\subsection{The Angular Correlation Function} \label{sect:acf}

\begin{figure}
\centering
\includegraphics[width=\columnwidth]{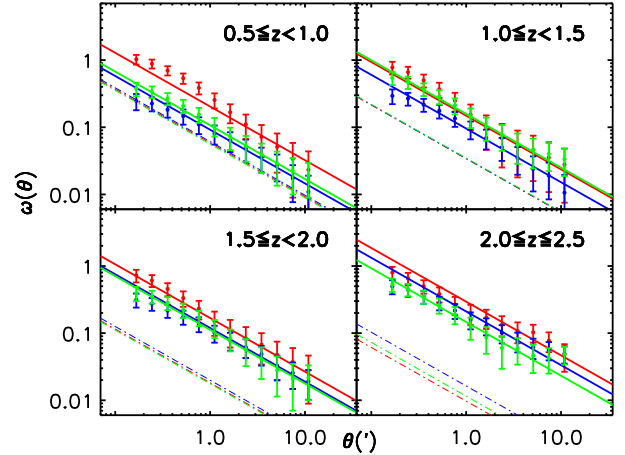}
\caption{The angular correlation functions of RS (red), GV (green), and BC (blue) galaxies with $0.5\le z\le 2.5$ and ${\rm log}(M_{\ast}/M_\odot)\ge10$. Integral constraint correction has been applied. The thick solid lines are the best-fit power-laws to the concerned sub-samples (see Section~\ref{sect:acf}), while the thin dot-dashed lines ({ almost overlapping with each other at lower redshift bins}) depict the angular correlation functions of DM halos hosting the relevant sub-samples in the corresponding redshift range (see Section~\ref{sect:bias}).}
\label{fig:ACF}
\end{figure}

The observed angular correlation functions of our galaxy samples are measured using the \citet{LS_1993} estimator:
\begin{equation}
\omega_{\rm obs}(\theta)=\frac{\rm DD(\theta)-2DR(\theta)+RR(\theta)}{\rm RR(\theta)}.
\end{equation}
At angular separations between 10 arcsec and 10 arcmin with the bin of $\Delta\log(\theta) = 0.16$, we measure the total numbers of the data-data (DD), data-random (DR), and random-random (RR) galaxy pairs, by generating 100,000 random galaxies and scattering them over the same survey area. These pair numbers are correspondingly normalized such that $\Sigma_{\theta}\rm DD(\theta)=\Sigma_{\theta}DR(\theta)=\Sigma_{\theta}RR(\theta)$. The size of the random sample is more than ten times larger than all our galaxy samples in any redshift bin. As shown in Figure~\ref{fig:ACF}, these initial estimates are further corrected for the so-called integral constraint, which is defined as
\begin{equation}
\omega_{\Omega}=\frac{1}{\Omega^{2}}\iint_{\Omega}\omega_{\rm T}(\theta_{12})d\Omega_{1}d\Omega_{2},
\end{equation}
where $\Omega$ is the total survey areas, $\theta_{12}$ is the angular separation of two arbitrary small solid angles $d\Omega_{1}$ and $d\Omega_{2}$, and $\omega_{\rm T}(\theta_{12})$ is the true angular correlation function of galaxy sample. In the limited survey field, the integral constraint significantly affects the clustering amplitude. To estimate the integral constraint, a power-law of $A_{\omega}\theta^{-\beta}$ is assumed for the true angular correlation function, and then a preliminary clustering amplitude, $A_{\omega}$, can be fit\footnote{http://purl.com/net/mpfit} in the log $\theta$ - log $\omega$ space by
\begin{equation}
\omega_{\rm obs}(\theta)=A_{\omega}\theta^{-\beta} - \omega_{\Omega}(A_\omega),
\end{equation}
where $\beta$ is fixed to 0.8. At very small angular separations, some of our angular correlation functions deviate from a single power-law, which is probably due to internal properties. In order to avoid this effect, we fit our angular correlation functions at the scales of 1 arcmin to 10 arcmin for all our sub-samples. In turn, the uncertainties of the corrected angular correlation functions are derived using the covariance matrix described in APPENDIX A of \citet{Brown_2008}, assuming the above best-fit true angular correlation function. Finally, for each sample, its clustering amplitude is refitted again by a power-law with the same index but considering the uncertainties. The ultimate fits are illustrated as the solid lines in Figure~\ref{fig:ACF}, and $A_{\omega}$ with 1 $\sigma$ uncertainty are tabulated in Table~\ref{tab:galaxy_bin}.

The selected value of $\beta$ is consistent with most previous observations \citep{Foucaud_2010,Furusawa_2011,Hartley_2013,Palamara_2013,Lin_2012}. We should stress that $\beta=0.8$ is not always true considering the one-halo term that two or more galaxies may reside in a common DM halo \citep{Zehavi_2004}. Since we mainly focus on massive galaxies with $\log(M_{\ast}/M_\odot)\ge 10$, we have made a simple assumption that there is only one galaxy residing in one DM halo. Note that if we free $\beta$, the best-fit $\beta$ is still close to 0.8 for our sub-samples in each redshift bin.

Figure~\ref{fig:ACF} shows the corrected angular correlation functions of our sub-samples with $\log(M_{\ast}/M_\odot)\ge 10$. An increasing trend of clustering amplitude, $A_{\omega}$, that runs through BC, GV toward RS galaxies, is observed in all redshift bins.

To explore the effects of mass completeness on the clustering, we also fit the angular correlation functions of sub-samples with $\log(M_{\ast}/M_\odot)\ge 10.5$ and $\log(M_{\ast}/M_\odot)\ge 11$. However, we do not illustrate the resultant angular correlation functions, since their shapes show no difference from those of sub-samples with $\log(M_{\ast}/M_\odot)\ge 10$, then the information they carry is somewhat redundant.

Since our sub-samples spread over a relatively broad redshift range, it is necessary to explore if there is any potential evolution in each redshift bin. We then divide our initial samples with $\log(M_{\ast}/M_\odot)\ge 10$ into smaller redshift bins of $\Delta z=0.1$ for $z=0.5-1.5$ and $\Delta z=0.25$ for $z>1.5$, ensuring each sub-sample contains more than two hundred galaxies. For each sub-sample, the evolution of its angular correlation function inside each redshift bin is generally weak and fluctuates around the angular correlation function of the initial sample constructed with larger redshift bin of $\Delta z=0.5$. The fluctuations are relatively small at $z\le1.5$, but larger at $z>1.5$ due to the smaller sample size at higher redshift. For this reason, in the following, we do not further discuss the evolution of the angular correlation function of our sub-samples inside each redshift bin.

\subsection{The Spatial Clustering and Correlation Length} \label{sect:sc}

\begin{figure}
\centering
\includegraphics[width=\columnwidth, angle=0]{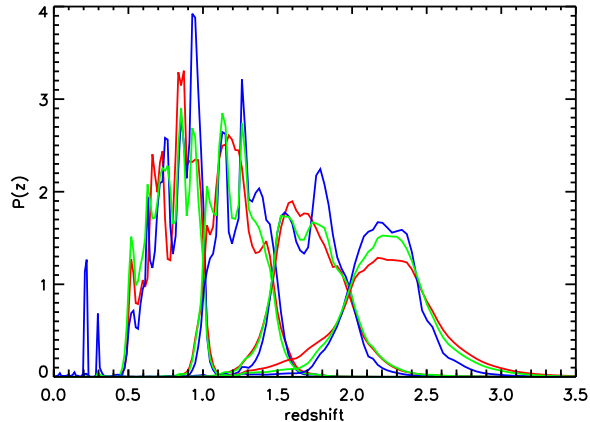}
\caption{The redshift probability distribution functions (PDFs) of RS (red), GV (green), and BC (blue) galaxies in different redshift bins selected by $\log(M_\ast/M_\odot)\ge 10$.
They are derived by averaging the redshift PDFs outputted by EAZY for all sources in the corresponding sub-samples.}
\label{fig:PDFz}
\end{figure}

The spatial correlation function, $\xi(r, z)$, is related to $\omega_{\rm T}(\theta)$ by the \citet{Limber_1954} equation:
\begin{equation}
\omega_{\rm T}(\theta)=\frac{\int_0^\infty \{\int_0^\infty\xi[r(\theta,z,z'), \bar z]\frac{dN}{dz'}dz'\} \frac{dN}{dz} dz}{(\int_0^{\infty}\frac{dN}{dz} dz)^2},
\end{equation}
where $\bar z$ is the mean redshift of galaxies at $z$ and $z'$, $r(\theta,z,z')$ is the spatial distance between galaxies at $z$ and $z'$ with an angular separation $\theta$, and $dN/dz$ is the redshift distribution of each galaxy population (cf. Figure~\ref{fig:PDFz}). Assuming a power-law, i.e., $\xi(r, z)=[{r}/{r_{0}(z)}]^{-\gamma}$, where $r_{0}(z)$ is the (comoving) spatial correlation length measured at $z$, and $\gamma$ is the slope of the power-law. For the spatial correlation function under the Limber and small angle approximation, one would end up with \citep[e.g.,][]{Overzier_2003,Quadri_2007,Hickox_2011,Toba_2017}
\begin{equation}
A_{\omega} = H_{\gamma} \frac{\int_0^\infty r^{\gamma}_{0}(z) [r_{c}(z)]^{1-\gamma} \left(\frac{dN}{dz}\right)^2 E(z)dz}{(c/H_{0})[\int_0^\infty \frac{dN}{dz} dz]^2},
\end{equation}
where $H_{\gamma} = \Gamma({1}/{2}) {\Gamma[(\gamma-1)/2]}/{\Gamma(\gamma/2)}$, $E(z)=\sqrt{\Omega_{m}(1+z)^{3}+\Omega_{\Lambda}}$, and the comoving radial distance $r_c(z) = (c/H_0) \int_0^z dz/E(z)$. Note that the slope $\gamma$ is related to $\beta$ by $\gamma = \beta + 1$. The evolution of spatial correlation with redshift, $r_0(z)$, is often linked to $r_0$ at $z = 0$ by $r_0(z) = r_0 (1+z)^{1-(3+\epsilon)/\gamma}$ with a specific ``clustering index'' $\epsilon$. Here, $\epsilon=\gamma-3$ is for ``constant clustering'', $\epsilon=0$ is for ``stable clustering'', and $\epsilon=\gamma-1$ is for the case of ``linear growth'' \citep[e.g.,][]{Blake_2002,Overzier_2003}. Referring to Figure~\ref{fig:PDFz}, since the redshift distributions of our sub-samples are relatively narrow, using different value of $\epsilon$ would not significantly change our results. Therefore, we assume $\epsilon=\gamma-3$ for a constant clustering over each redshift bin and then assess any possible redshift evolution of $r_0$ among redshift bins for each galaxy population (see Figure~\ref{fig:r0_all} and Table~\ref{tab:galaxy_bin} for the estimations of $r_0(\bar z)$ for each galaxy population in each redshift bin and Section~\ref{sect:evo_r0} for discussions). Note that the uncertainty of $r_0$ is directly propagated from that of $A_\omega$.

\subsection{Bias and Dark Matter Halo Mass} \label{sect:bias}

To assess the corresponding DM halo mass of a galaxy population in a redshift bin, we should firstly derive the bias factor, $b_{\rm gal}$, for that galaxy population, which relates the spatial correlation function of galaxies, $\xi_{\rm gal}$, to that of the underlying DM halos, $\xi_{\rm DM}$, through $\xi_{\rm gal}=b^2_{\rm gal} \xi_{\rm DM}$ \citep{Myers_2006}. Although the bias factor might be related to the properties of halo assembly \citep{Xu_2017} and halo formation histories \citep{Desjacques_2016} at large scales, it is often assumed to be a scale-independent constant over each redshift bin. After projection, we have $b_{\rm gal} = \sqrt{\omega_{\rm gal} / \omega_{\rm DM}}$, where $\omega_{\rm gal} \sim A_\omega^{\rm gal}$ is the observed angular correlation function of the concerned galaxy population.

The angular correlation function of DM halos, $\omega_{\rm DM}(\theta)$, hosting the concerned galaxy population with redshift distribution of $dN/dz$ is computed following \citet{Myers_2007}. We firstly generalize a power spectrum of the underlying DM, $\Delta_{\rm NL}(k,z)$, using HALOFIT \citep{Smith_2003}. Then the spatial correlation function derived from $\Delta_{\rm NL}(k,z)$ can be transformed to the angular correlation function $\omega_{\rm DM}(\theta)$ by
\begin{equation}
\begin{split}
\omega_{\rm DM}(\theta)=\frac{H_0\pi}{c}\int_{z=0}^\infty\int_{k=0}^\infty{\frac{\Delta_{\rm NL}^2(k,z)}{k}J_0[k\theta\chi(z)](\frac{dN}{dz})^2}\\
{\sqrt{\Omega_m(1+z)^3+\Omega_\Lambda}F(\chi)\frac{dk}{k}dz},
\end{split}
\end{equation}
where $J_0$ is the first kind zeroth-order Bessel function, $\chi$ is the radial comoving distance, and ${dN}/{dz}$ is the normalized redshift distribution of the relevant galaxy population so that $\int^\infty_0 (dN/dz) dz = 1$. A flat cosmology implies $F(\chi)=1$. Thereafter, a power-law of $A^{\rm DM}_\omega \theta^{-\beta}$ with $\beta = 0.8$ is fit to $\omega_{\rm DM}(\theta)$ in the same aforementioned range of $\theta$. For the three galaxy populations in four redshift bins, these best-fit power-laws for DM halos are plotted as the thin dot-dashed lines in Figure~\ref{fig:ACF}.

The corresponding bias, $b_{\rm gal}$, for a galaxy population then can be calculated by the square root of the amplitude ratio of the angular correlation function of the galaxy population to that of the DM halo, i.e., $b_{\rm gal} = \sqrt{A_\omega^{\rm gal} / A^{\rm DM}_\omega}$ (see Table~\ref{tab:galaxy_bin}). The uncertainty of bias is again propagated from that of $A_\omega^{\rm gal}$ and we have neglected the uncertainty of $A^{\rm DM}_\omega$ induced by the clustering of DM halos.

Since the typical mass of DM halo hosting a galaxy population is related to the bias of that galaxy population, following \citet{Sheth_2001}, we approximately convert the bias of each galaxy population to the corresponding host halo mass, $M_{\rm halo}$ (see Table~\ref{tab:galaxy_bin}), while the related discussions are delayed to Section~\ref{sect:evo_halomass}.

\subsection{Modeling Clustering of Galaxies} \label{sect:clustering_model}

In order to compare with the observed spatial clustering of galaxies, we model the evolution of spatial clustering of galaxies hosted within DM halos with a constant or growing mass. For DM halos with mass of $M_{\rm halo}$ at redshift $z$, we firstly fit a power-law of $\xi_{\rm DM}(r,z) = [r/r_0(z)]^{-\gamma}$ with $\gamma = 1.8$ to the spatial correlation function of DM halo as introduced in the last section and then multiplied by the halo bias, $b^2(M_{\rm halo}, z)$, to approximate the spatial correlation function of galaxies hosted within the given DM halos.

For a growing halo, we assume that the mass of DM halo increases at a median growth rate of
\begin{equation}
\begin{split}
<\dot{M}>=25.3~\left(\frac{M_{\rm halo}}{10^{12}~M_{\odot}}\right)^{1.1} (1+1.65z)\\
\times\sqrt{\Omega_m(1+z)^3+\Omega_{\Lambda}}~~{M_{\odot}~\rm yr^{-1}},
\end{split}
\end{equation}
where $M_{\rm halo}$ is the halo mass at redshift $z$ \citep{Fakhouri_2010}.

\begin{table*} \caption{Best-fit clustering properties of galaxy sub-samples in COSMOS/UltraVISTA field\label{tbl-1}}
\centering

\begin{tabular}{lcccccccccc}
\hline
\hline
Galaxy sub-sample &
\(N_{\rm source}^{\rm a}\) &
$\bar{z}^{\rm b}_{\rm phot}$ &
$\log(M_\ast/M_\odot)^{\rm c}$ &
$A^{\rm d}_\omega$ &
\(r_{0}/[h^{-1}{\rm Mpc}]^{\rm e}\) &
\(b_{\rm gal}^{\rm f}\) &
$\log(M_{\rm halo}/[h^{-1}{M_\odot}])^{\rm g}$ \\
\hline
$0.5 \le z < 1.0$ & \multicolumn{7}{l}{ $\log(M_\ast/M_\odot) > 10$ \& 80\% mass-completeness } &  \\
$\rm RS$ & 6051 & 0.823 & 10.53 & 0.204$\pm$0.034 & \(6.88_{-0.66}^{+0.62}\) & 1.87$\pm$0.16 & \(12.98_{-0.16}^{+0.14}\)   \\
$\rm GV$ & 4116 & 0.789 & 10.39 & 0.107$\pm$0.013 & \(5.00_{-0.35}^{+0.33}\) & 1.38$\pm$0.08 & \(12.40_{-0.15}^{+0.13}\)    \\
$\rm BC$ & 3774 & 0.843 & 10.21 & 0.092$\pm$0.011 & \(4.24_{-0.29}^{+0.27}\) & 1.23$\pm$0.07 & \(12.03_{-0.18}^{+0.15}\)    \\
 & \multicolumn{7}{c}{ $\log(M_\ast/M_\odot) > 10.5$ \& 98\% mass-completeness } &  \\
$\rm RS$ & 3234 & 0.819 & 10.80 & 0.201$\pm$0.034 & \(6.66_{-0.65}^{+0.60}\) & 1.80$\pm$0.15 & \(12.92_{-0.17}^{+0.14}\)   \\
$\rm GV$ & 1546 & 0.800 & 10.67 & 0.115$\pm$0.020 & \(5.07_{-0.51}^{+0.47}\) & 1.41$\pm$0.12 & \(12.43_{-0.22}^{+0.19}\)    \\
$\rm BC$ & 537 & 0.856 & 10.59 & 0.123$\pm$0.030 & \(4.73_{-0.68}^{+0.60}\) & 1.36$\pm$0.17 & \(12.28_{-0.37}^{+0.27}\)    \\
 & \multicolumn{7}{r}{ $\log(M_\ast/M_\odot) > 11$ \& 100\% mass-completeness } &  \\
$\rm RS$ & 855 & 0.820 & 11.16 & 0.289$\pm$0.058 & \(7.95_{-0.93}^{+0.85}\) & 2.11$\pm$0.21 & \(13.19_{-0.18}^{+0.15}\)   \\
\hline
$1.0 \le z < 1.5$ & \multicolumn{7}{l}{ $\log(M_\ast/M_\odot) > 10$ \& 80\% mass-completeness } & \\
$\rm RS$ & 5099 & 1.223 & 10.51 & 0.149$\pm$0.031 & \(6.30_{-0.77}^{+0.70}\) & 2.09$\pm$0.22 & \(12.76_{-0.20}^{+0.17}\)    \\
$\rm GV$ & 2143 & 1.219 & 10.44 & 0.158$\pm$0.022 & \(6.46_{-0.52}^{+0.48}\) & 2.13$\pm$0.15 & \(12.80_{-0.13}^{+0.12}\)    \\
$\rm BC$ & 5798 & 1.270 & 10.26 & 0.096$\pm$0.011 & \(4.85_{-0.32}^{+0.30}\) & 1.67$\pm$0.10 & \(12.27_{-0.13}^{+0.12}\)    \\
 & \multicolumn{7}{c}{ $\log(M_\ast/M_\odot) > 10.5$ \& 98\% mass-completeness } &  \\
$\rm RS$ & 2602 & 1.222 & 10.76 & 0.188$\pm$0.043 & \(7.17_{-0.96}^{+0.87}\) & 2.29$\pm$0.26 & \(12.92_{-0.21}^{+0.18}\)    \\
$\rm GV$ & 921 & 1.216 & 10.69 & 0.176$\pm$0.025 & \(6.85_{-0.56}^{+0.52}\) & 2.20$\pm$0.16 & \(12.85_{-0.13}^{+0.12}\)    \\
$\rm BC$ & 1192 & 1.297 & 10.63 & 0.183$\pm$0.026 & \(6.95_{-0.57}^{+0.53}\) & 2.26$\pm$0.16 & \(12.85_{-0.13}^{+0.11}\)    \\
 & \multicolumn{7}{r}{ $\log(M_\ast/M_\odot) > 11$ \& 100\% mass-completeness } &  \\
$\rm RS$ & 494 & 1.217 & 11.13 & 0.241$\pm$0.069 & \(8.23_{-1.41}^{+1.23}\) & 2.47$\pm$0.35 & \(13.04_{-0.26}^{+0.20}\)    \\
\hline
$1.5 \le z < 2.0$ & \multicolumn{7}{l}{ $\log(M_\ast/M_\odot) > 10$ \& 80\% mass-completeness } & \\
$\rm RS$ & 3354 & 1.729 & 10.50 & 0.168$\pm$0.032 & \(7.71_{-0.85}^{+0.78}\) & 3.04$\pm$0.29 & \(12.91_{-0.16}^{+0.14}\)    \\
$\rm GV$ & 1393 & 1.727 & 10.38 & 0.112$\pm$0.016 & \(6.24_{-0.51}^{+0.48}\) & 2.52$\pm$0.18 & \(12.61_{-0.13}^{+0.12}\)    \\
$\rm BC$ & 4971 & 1.733 & 10.29 & 0.119$\pm$0.014 & \(5.99_{-0.40}^{+0.38}\) & 2.44$\pm$0.14 & \(12.55_{-0.11}^{+0.10}\)    \\
 & \multicolumn{7}{c}{ $\log(M_\ast/M_\odot) > 10.5$ \& 98\% mass-completeness } &  \\
$\rm RS$ & 1719 & 1.735 & 10.74 & 0.221$\pm$0.053 & \(8.98_{-1.27}^{+1.14}\) & 3.27$\pm$0.39 & \(13.02_{-0.20}^{+0.16}\)    \\
$\rm GV$ & 506 & 1.735 & 10.69 & 0.151$\pm$0.037 & \(7.37_{-1.06}^{+0.95}\) & 2.73$\pm$0.33 & \(12.74_{-0.22}^{+0.18}\)    \\
$\rm BC$ & 1316 & 1.780 & 10.67 & 0.144$\pm$0.015 & \(6.65_{-0.39}^{+0.38}\) & 2.58$\pm$0.13 & \(12.64_{-0.09}^{+0.09}\)    \\
 & \multicolumn{7}{r}{ $\log(M_\ast/M_\odot) > 11$ \& 100\% mass-completeness } &  \\
$\rm RS$ & 250 & 1.729 & 11.12 & 0.601$\pm$0.260 & \(15.66_{-4.23}^{+3.46}\) & 5.13$\pm$1.11 & \(13.62_{-0.31}^{+0.23}\)    \\
\hline
$2.0 \le z \le 2.5$ & \multicolumn{7}{l}{ $\log(M_\ast/M_\odot) > 10$ \& 80\% mass-completeness } &  \\
$\rm RS$ & 1251 & 2.231 & 10.54 & 0.294$\pm$0.042 & \(12.21_{-1.00}^{+0.94}\) & 5.50$\pm$0.39 & \(13.36_{-0.10}^{+0.09}\)   \\
$\rm GV$ & 1029 & 2.243 & 10.46 & 0.147$\pm$0.022 & \(7.56_{-0.65}^{+0.61}\) & 3.57$\pm$0.27 & \(12.75_{-0.12}^{+0.11}\)   \\
$\rm BC$ & 2239 & 2.215 & 10.31 & 0.211$\pm$0.025 & \(8.57_{-0.58}^{+0.55}\) & 3.59$\pm$0.21 & \(12.78_{-0.10}^{+0.09}\)   \\
 & \multicolumn{7}{c}{ $\log(M_\ast/M_\odot) > 10.5$ \& 98\% mass-completeness } &  \\
$\rm RS$ & 697 & 2.264 & 10.71 & 0.374$\pm$0.084 & \(13.96_{-1.84}^{+1.66}\) & 5.67$\pm$0.64 & \(13.40_{-0.15}^{+0.13}\)   \\
$\rm GV$ & 477 & 2.224 & 10.69 & 0.286$\pm$0.057 & \(10.94_{-1.27}^{+1.16}\) & 4.90$\pm$0.49 & \(13.21_{-0.14}^{+0.12}\)   \\
$\rm BC$ & 662 & 2.194 & 10.70 & 0.259$\pm$0.040 & \(9.60_{-0.85}^{+0.80}\) & 4.38$\pm$0.34 & \(13.07_{-0.11}^{+0.10}\)   \\
\hline
\end{tabular}

%Notes: \\
\begin{tabular}{p{17cm}}
$^{\rm a}$ The source number of each galaxy sub-sample. \\
$^{\rm b}$ The median photometric redshift of each galaxy sub-sample. \\
$^{\rm c}$ The median stellar mass of each galaxy sub-sample. \\
$^{\rm d}$ The clustering amplitude of the angular correlation function of each galaxy sub-sample (see Section~\ref{sect:acf} and Figure~\ref{fig:ACF}). \\
$^{\rm e}$ The correlation length of the spatial correlation function of each galaxy sub-sample (see Sections~\ref{sect:sc} and \ref{sect:evo_r0}, and Figures~\ref{fig:r0_all} and \ref{fig:r0_compare}). \\
$^{\rm f}$ The bias of each galaxy sub-sample (see Section~\ref{sect:bias}). \\
$^{\rm g}$ The halo mass approximately converted from the bias (see Sections~\ref{sect:bias} and \ref{sect:evo_halomass}, and Figure~\ref{fig:halomass_all}). \\
\end{tabular}

\label{tab:galaxy_bin}
\end{table*}

\section{Results and Discussions} \label{sect:RaD}

Our three galaxy populations, i.e., the RS, GV, and BC galaxies, are selected by stellar masses of $\log(M_{\ast}/M_{\odot}) \ge 10$ at $0.5\le z\le2.5$ and separated by the rest-frame extinction-corrected UV colors \citep[cf. Section~\ref{sect:sample_construction} and ][]{Wang_2017}. Using the recipes described in Section~\ref{sect:clustering}, we discuss the evolutions of correlation length and DM halo mass of RS, GV, and BC sub-samples in Sections~\ref{sect:evo_r0} and \ref{sect:evo_halomass}, respectively, while the effect of growing stellar mass on the clustering evolution in Section~\ref{sect:acc_evo}. Finally, the role of GV galaxies played in massive galaxy evolution and its relationship to AGN activity are discussed according to the clustering similarity between GV galaxies and AGNs in Section~\ref{sect:GV_clustering}.

\subsection{The Evolution of Correlation Length}\label{sect:evo_r0}

\begin{figure}
\centering
\includegraphics[width=\columnwidth]{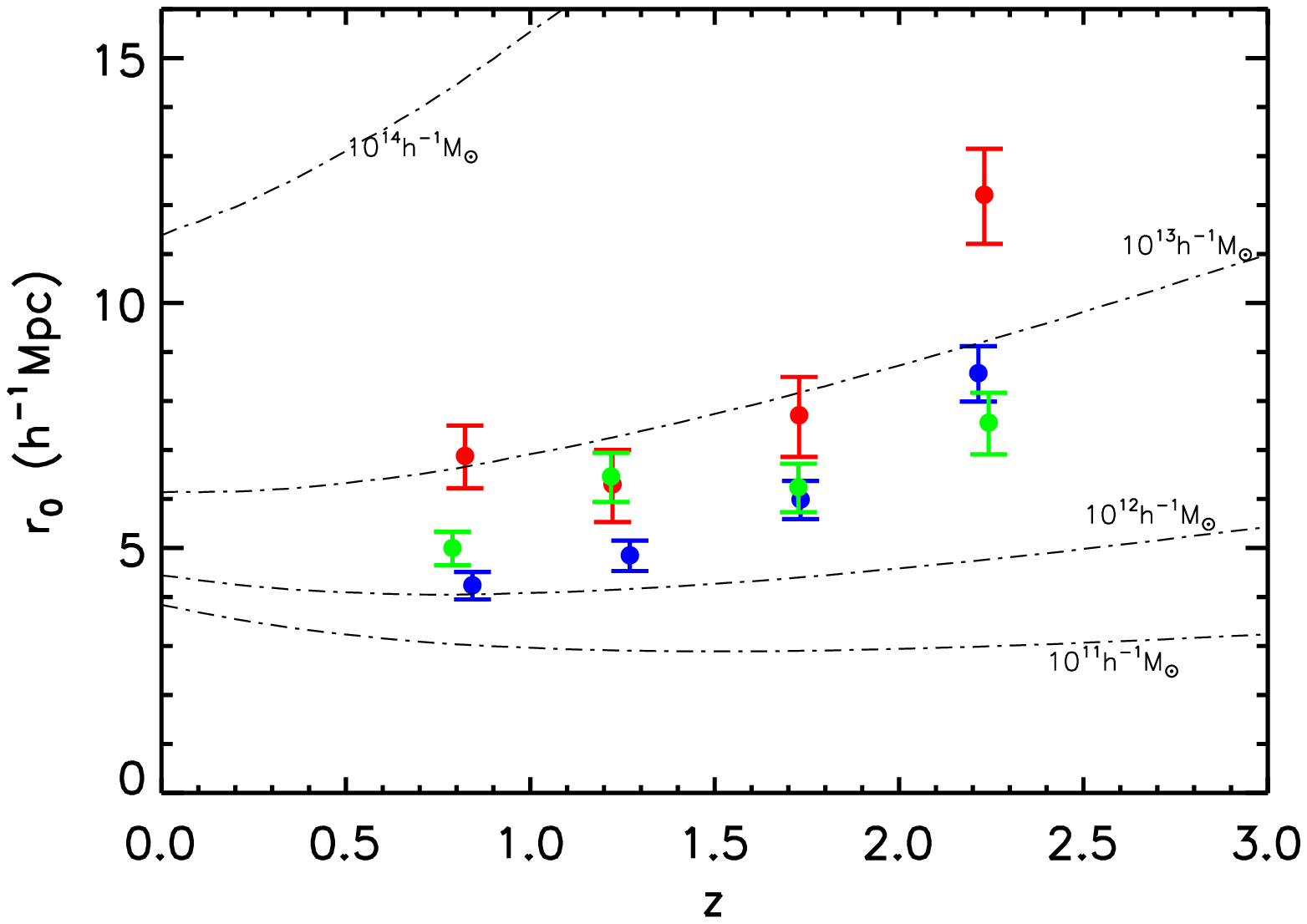}
\caption{The correlation lengths, \(r_{0}\), as a function of redshift for the massive RS (red), GV (green), and BC (blue) galaxies, selected with $\log(M_{\ast}/M_\odot) \ge 10$. The thin dot-dashed lines show the evolutions of $r_0$ for given DM halos with constant masses ranging from $10^{11}$ to $10^{14}~{h}^{-1}M_\odot$.}
\label{fig:r0_all}
\end{figure}

\begin{figure}
\centering
\includegraphics[width=\columnwidth]{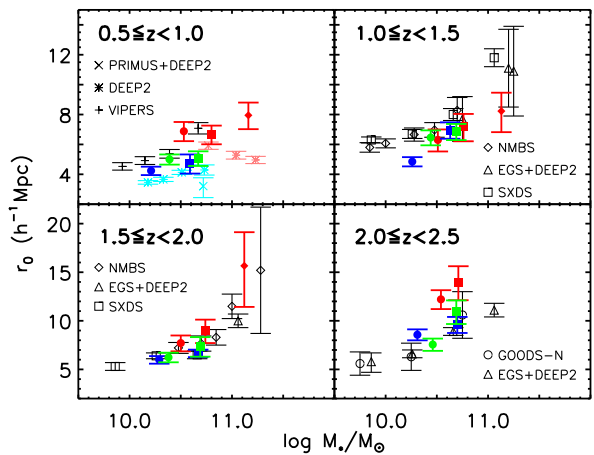}
\caption{The correlation lengths, $r_0$, in four redshift bins as a function of stellar mass for the RS (red), GV (green), and BC (blue) galaxy sub-samples, selected with $\log(M_{\ast}/M_\odot) \ge 10$ (filled circles), 10.5 (filled squares), and 11 (filled diamonds), and compared with various galaxy samples from the literature, including results of the VIMOS Public Extragalactic Redshift Survey by \citet[VIPERS,][pluses]{Marulli_2013}, the DEEP2 survey by \citet[][stars]{Mostek_2013}, the PRIMUS+DEEP2 survey by \citet[][crosses]{Coil_2017}, the NMBS survey by \citet[][open diamonds]{Wake_2011}, the EGS+DEEP2 survey by \citet[][open triangles]{Foucaud_2010}, the Subaru/XMM-Newton Deep Survey (SXDS) by \citet[][open squares]{Furusawa_2011}, and the GOODS-N field by \citet[][open circles]{Lin_2012}.}
\label{fig:r0_compare}
\end{figure}

The correlation lengths, \(r_{0}\), as a function of redshift for the massive RS, GV, and BC sub-samples, selected with $\log(M_{\ast}/M_\odot) \ge 10$, are shown in Figure~\ref{fig:r0_all} and are tabulated in Table~\ref{tab:galaxy_bin}. In general, the correlation lengths of all galaxy sub-samples increase with increasing redshift. The correlation lengths of RS galaxies are larger than those of BC galaxies in all given redshift bins, while the correlation lengths of GV galaxies generally lie between them. These reflect the fact that the clustering amplitude not only depends on the stellar mass of galaxies but also on their color or SFR \citep{Kim_2015,Lin_2012}.

In Figure~\ref{fig:r0_all}, the correlation lengths of DM halos with constant masses are compared (see Section~\ref{sect:clustering_model}). The correlation lengths of the RS galaxies are found to generally follow those of a constant DM halo with mass of $10^{13}~M_\sun$. Since we use an identical stellar mass threshold to select our RS galaxies in different redshift bins, this does not reflect the clustering evolution of a certain galaxy population, but the fact that originally less massive halos join the analyzed samples at lower redshift ranges. Galaxies of the same stellar mass at different redshifts are not expected to be the same galaxies through cosmic time. At $z>2$ there are large (strongly clustered) galaxies, while today they would be pretty average. On the other hand, the RS galaxies here seem to follow roughly the same halos at least from $z\le2$, which may suggest that by $z\sim2$ the RS galaxies have assembled their stellar masses and did not evolve much since then. This is not the case of the GV and BC galaxies which clearly belong to different halo populations at different redshifts.

Note that although we have used the same stellar mass threshold to select the RS, GV, and BC sub-samples, the median stellar mass of GV sub-samples is generally smaller/larger than that of RS/BC sub-samples (see Table~\ref{tab:galaxy_bin}). Therefore, more massive galaxies generally have larger correlation lengths, which is better illustrated in Figure~\ref{fig:r0_compare} for the correlation lengths of different galaxy sub-samples as a function of stellar mass in four redshift bins. This may mean that the dependence of clustering on galaxy color is actually an indirect result of its dependence on stellar mass (see Section~\ref{sect:acc_evo} for further discussion).

In Figure~\ref{fig:r0_compare}, we also compare our correlation lengths of RS, GV, and BC sub-samples, selected with different stellar mass thresholds, to those selected with similar stellar masses and in similar redshift intervals from the literature. Generally speaking, our results globally agree with previous works, but in the lowest redshift bin, our results show a stronger clustering than others. This may be due to the presence of several large structures at $z<1$ in the COSMOS field \citep{Mendez_2016}. Some other studies \citep[e.g.,][]{Coil_2008,Hickox_2009,Zehavi_2011} use a magnitude/luminosity cut to select galaxies for clustering analyses. Since the conversion between magnitude and stellar mass is not so apparent, we do not show them in Figure~\ref{fig:r0_compare}.

We also find consistency when comparing our correlation lengths as a function of redshift or stellar mass with those of the recent IllustrisTNG simulation \citep[][see their Figure 13]{Springel_2017}, who studied the correlation functions of simulated galaxies selected in different stellar mass ranges and found that the galactic clustering length depends strongly on stellar mass and redshift.

\subsection{The Evolution of Halo Mass}\label{sect:evo_halomass}

\begin{figure}
\centering
\includegraphics[width=\columnwidth, angle=0]{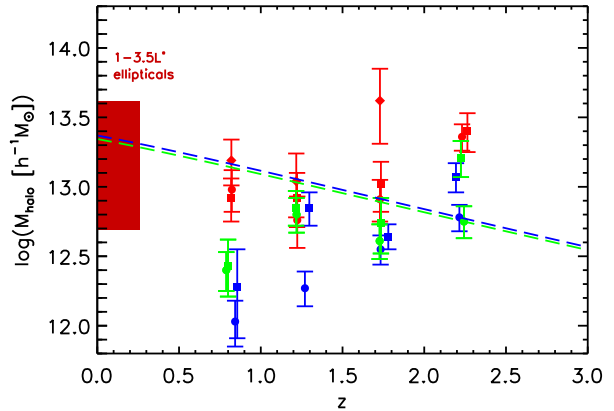}
\caption{The evolution of halo mass versus redshift for the massive RS (red), GV (green), and BC (blue) galaxy sub-samples, selected with $\log(M_{\ast}/M_\odot) \ge $ 10 (circles), 10.5 (squares), and 11 (diamonds).
The red region indicates the halo masses of local luminous early-type galaxies with $r$-band luminosities of 1-3.5 $L^*$ \citep{Zehavi_2011}.
The blue/green thick dashed line shows the growth of halo mass hosting typical BC/GV galaxies at $z \sim 2.2$ (see Section~\ref{sect:clustering_model}).
}
\label{fig:halomass_all}
\end{figure}

In Figure~\ref{fig:halomass_all} we show the evolution of halo mass of our galaxy sub-samples selected  with $\log(M_{\ast}/M_\odot) \ge $ 10, 10.5, and 11. The typical halo mass of RS galaxies roughly stays unchanged along ${\rm log}(M_{\rm halo}/[h^{-1}{M_{\odot}}]) = 13$ across redshifts at least from $z\sim2$. Since we adopt a uniform stellar mass threshold to select all massive galaxies within each redshift bin, this does not reflect the clustering evolution of a certain galaxy population. A possible mechanism accounting for this is that the originally less massive halos join more massive halos and lower the clustering amplitude of RS galaxies toward lower redshifts, making it roughly constant. Those sources within less massive halos are likely to be GV and BC galaxies at higher redshifts. These galaxies truncate their star formations and transit their colors redward. On the other hand, the RS galaxies may not accumulate much stellar mass since $z \sim 2$.

At all redshifts up to $z \sim 2$, the selection effect has a limited impact on the halo mass of RS galaxies, unless the mass threshold has been increased up to $\log(M_\ast/M_\odot) \ge 11$. This result is consistent with that of \citet{Zehavi_2011}, who found that, at $z < 0.3$, over a large luminosity range, the clustering of red galaxies evolves little, except when the luminosity is lifted up to $L>4L_\ast$.

The typical halo masses of BC and GV galaxies generally decrease with decreasing redshift. This is consistent with the downsizing picture that massive objects form first. The halo mass of GV galaxies generally lies between those of RS and BC galaxies as expected, and slightly fluctuates around a mean halo mass of $\log (M_{\rm halo}/[h^{-1}{M_{\odot}}]) = 12.7$ across redshifts.

Nevertheless, the halo mass of GV galaxies shows slightly different tendencies at higher and lower redshift. At higher redshifts (i.e., $z > 1.5$), the clustering amplitudes and halo masses of GV galaxies are all closer to those of BC galaxies, which suggests that they may be born in similar halos. \citet{Barro_2015} found that some internal physical mechanisms, like compaction, may decrease the star formation of galaxies at high redshift. This may be responsible for the transition of BC to GV galaxies at high redshift without changing their clustering properties. At lower redshifts (i.e., $z < 1.5$), the clustering amplitudes and halo masses of GV galaxies are instead closer to those of RS galaxies. Some external physical processes, such as gas rich mergers \citep[e.g.,][]{Fakhouri_2010,Barro_2013,Barro_2014,Toft_2014}, may have significantly increased the halo mass of GV galaxies and convert their environment to that similar to that of RS galaxies at lower redshift. These results are consistent with \citet{Pan_2013}, who found that the environment where GV galaxies reside is closer to that of BC galaxies at higher redshifts but to RS galaxies at lower redshifts.

Also shown as the dashed lines in Figure~\ref{fig:halomass_all}, we model the evolutionary tracks of BC and GV galaxies at $z\sim2.2$ hosted within DM halos growing at a median growth rate given by \citet[][see Section~\ref{sect:clustering_model} for detail]{Fakhouri_2010}. The RS galaxies at lower redshifts as well as the local bright ellipticals lie just near this evolutionary sequence. It suggests that the GV/BC galaxies at $z \sim 2.2$ may be the progenitors of RS galaxies at $z < 1.5$ and the local massive early-type galaxies.

\subsection{The Effect of Stellar Mass Evolution on Clustering}\label{sect:acc_evo}

The stellar masses of BC/GV galaxies would continuously increase as long as their star formation proceeds, and after their star formation is truncated, they would become part of RS galaxy population at lower redshifts. Therefore, using a single stellar mass threshold to select galaxies with the same color but at different redshifts would not result in a self-consistent population across redshifts. Then, it is of interest to explore how the stellar mass evolution of a certain population affects their clustering evolution.

\begin{figure}
   \centering
   \includegraphics[width=\columnwidth, angle=0]{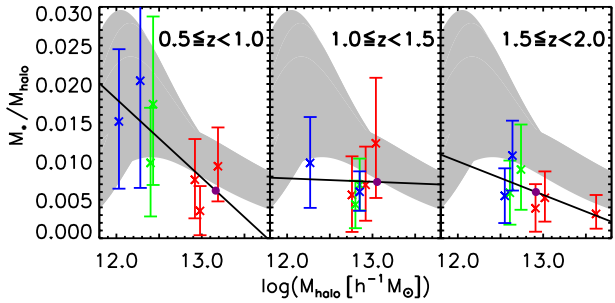}
   \caption{The stellar to halo mass ratio (SHMR) of the RS (red), GV (green), and BC (blue) galaxy sub-samples at $z \sim 0.8$ (crosses), $z \sim 1.2$ (triangles), and $z \sim 1.7$ (squares), evolved from the BC galaxies at $z\sim2.2$ (blue open circle; see Section~\ref{sect:acc_evo}), compared to the possible SHMRs simulated by \citet[][gray region]{Moster_2010}.
	}
   \label{fig:SHMR}
\end{figure}

Instead of using the same stellar mass threshold to select galaxies in different redshift bins, we use a redshift-dependent stellar mass threshold to select galaxies in the three lower redshift bins, i.e., $z \le 2$.
In the following, we take the BC galaxies at $z \sim 2.2$ as reference, but our results would not be much altered adopting the GV galaxies at $z \sim 2.2$.
Starting from the highest redshift bin of $2 \le z \le 2.5$ where the BC galaxies are selected with $\log(M_\ast/M_\odot) \ge 10$,
their host halos are assumed to grow up at a median growth rate given by \citet{Fakhouri_2010}, then the masses of halos hosting galaxies at lower redshifts evolved from the BC galaxies at $z \sim 2.2$ can then be derived. Using the median stellar and host halo masses tabulated in Table~\ref{tab:galaxy_bin}, we can estimate the stellar to halo mass ratios (SHMRs) for our galaxy sub-samples in each of the three lower redshift bins as shown in Figure~\ref{fig:SHMR}, where the possible SHMRs simulated by \citet{Moster_2010} are presented for comparison. In each redshift bin, these SHMRs versus logarithmic halo masses are fit by a linear regression (black line in Figure~\ref{fig:SHMR}) and a typical SHMR is correspondingly estimated for the halo hosting galaxies evolved from the BC galaxies at $z \sim 2.2$. This redshift-dependent typical SHMR is then used to derive the median stellar mass for the halo hosting galaxies evolved from the BC galaxies at $z \sim 2.2$.
Finally, a specific stellar mass threshold for galaxies in a lower redshift bin is tuned such that its median stellar mass is the same as that evolved from the BC galaxies at $z \sim 2.2$.
Once the stellar mass threshold has been determined, the clustering properties can be obtained following what presented above.

\begin{figure}
   \centering
   \includegraphics[width=\columnwidth, angle=0]{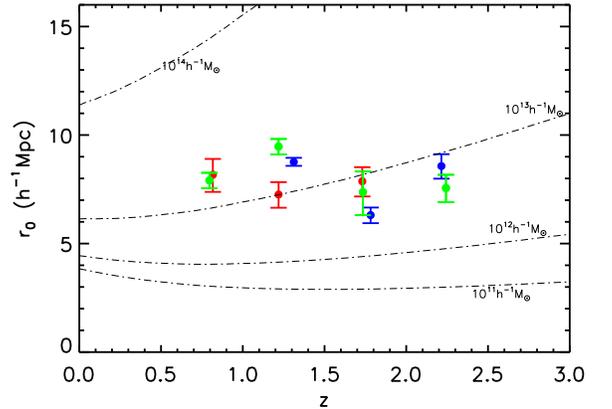}
   \caption{Same as Figure~\ref{fig:r0_all}, but for the correlation lengths of the RS (red), GV (green), and BC (blue) galaxy sub-samples, evolved from the BC galaxies at $z\sim2.2$ (see Section~\ref{sect:acc_evo}).
	}
   \label{fig:r0_evo}
\end{figure}

\begin{figure}
\centering
\includegraphics[width=\columnwidth, angle=0]{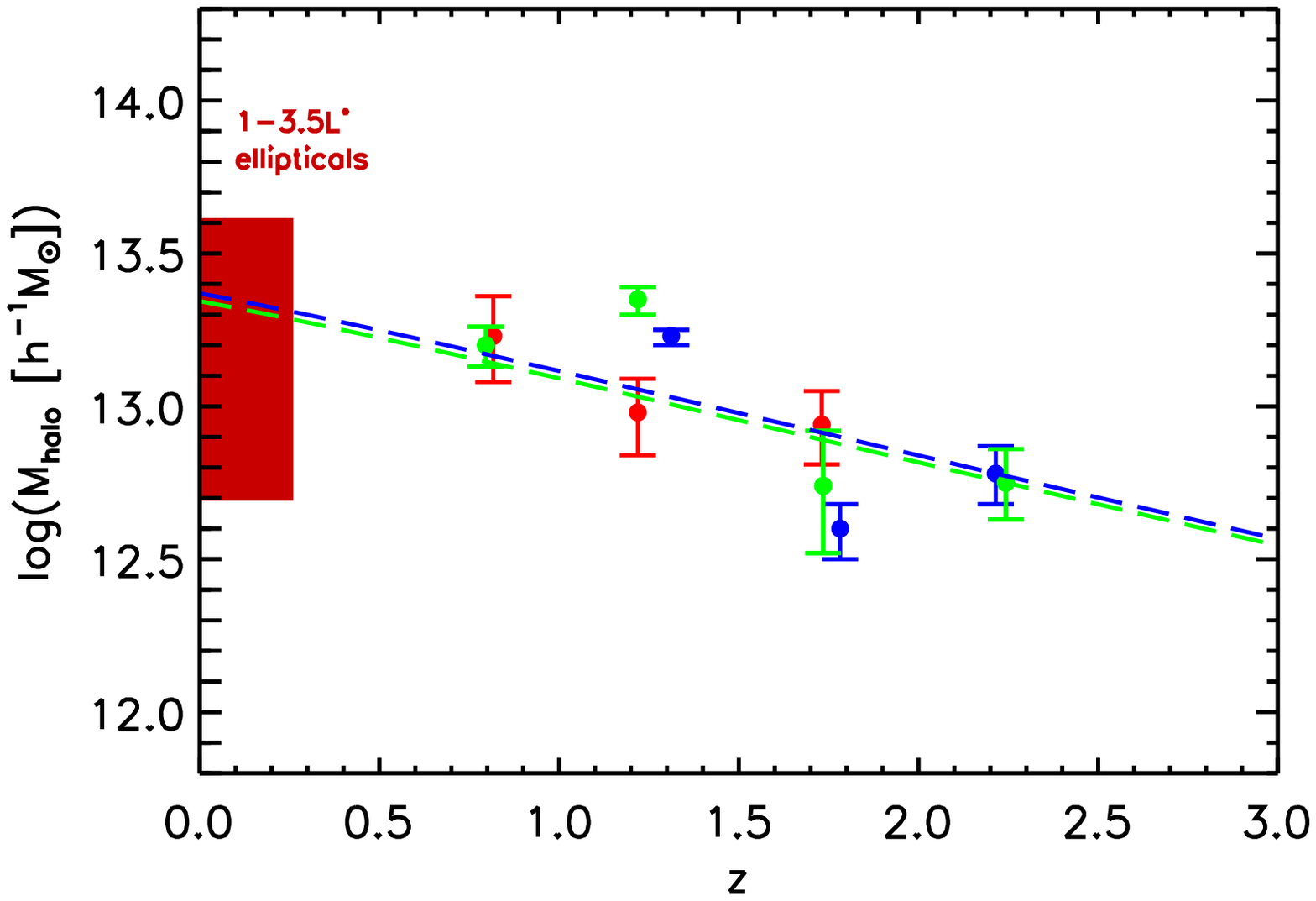}
\caption{Same as Figure~\ref{fig:halomass_all}, but for the halo masses of the RS (red), GV (green), and BC (blue) galaxy sub-samples, evolved from the BC galaxies at $z\sim2.2$ (see Section~\ref{sect:acc_evo}).
}
\label{fig:halomass_evo}
\end{figure}

The correlation lengths and halo masses of all galaxy sub-samples evolved from the BC galaxies at $z \sim 2.2$ are shown in Figures~\ref{fig:r0_evo} and \ref{fig:halomass_evo}, respectively, and are tabulated in Table~\ref{tbl-2}.
Compared to Figures~\ref{fig:r0_all} and \ref{fig:halomass_all}, the difference among the clustering properties of galaxies with different colors indeed decreases once their median stellar masses are the same.
One may have noticed from Table~\ref{tab:galaxy_bin} that the median stellar masses decrease with moving along the galaxy sequence from RS to GV to BC populations when a common stellar mass threshold is adopted for them. Therefore, the clustering differences as illustrated in Figures~\ref{fig:r0_all} and \ref{fig:halomass_all} may suggest that the median stellar mass alone is a good predictor of galaxy clustering, weakly independent on their colors.

\begin{table*} \caption{Best-fit clustering properties of galaxy sub-samples evolved from the BC galaxies at $z \sim 2.2$\label{tbl-2}}
\centering

\setlength{\tabcolsep}{0.8mm}{
\begin{tabular}{lcccccccccc}
\hline
\hline
Galaxy sub-sample &
\(N_{\rm source}^{\rm a}\) &
$\bar{z}^{\rm b}_{\rm phot}$ &
$\log(M_\ast/M_\odot)^{\rm c}_{\rm thresh}$ &
$\log(M_\ast/M_\odot)^{\rm d}$ &
$A^{\rm e}_\omega$ &
\(r_{0}/[h^{-1}{\rm Mpc}]^{\rm f}\) &
\(b_{\rm gal}^{\rm g}\) &
$\log(M_{\rm halo}/[h^{-1}{M_\odot}])^{\rm h}$ \\
\hline
$0.5 \le z < 1.0$ \\
$\rm RS$ & 2030 & 0.817 & 10.71 & 10.95 & 0.294$\pm$0.049 & \(8.17_{-0.79}^{+0.73}\) & 2.16$\pm$0.18 & \(13.23_{-0.15}^{+0.13}\)   \\
$\rm GV$ & 343 & 0.796 & 10.83 & 10.95 & 0.260$\pm$0.021 & \(7.91_{-0.36}^{+0.35}\) & 2.09$\pm$0.08 & \(13.20_{-0.07}^{+0.06}\)    \\
$\rm BC$ & 29 & $-$ & $-$ & $-$ & $-$ & $-$ & $-$ & $-$    \\
\hline
$1.0 \le z < 1.5$ \\
$\rm RS$ & 1448 & 1.219 & 10.72 & 10.92 & 0.211$\pm$0.031 & \(7.26_{-0.61}^{+0.57}\) & 2.38$\pm$0.17 & \(12.98_{-0.14}^{+0.11}\)    \\
$\rm GV$ & 225 & 1.219 & 10.80 & 10.92 & 0.335$\pm$0.023 & \(9.47_{-0.36}^{+0.35}\) & 3.03$\pm$0.10 & \(13.35_{-0.05}^{+0.04}\)    \\
$\rm BC$ & 151 & 1.312 & 10.84 & 10.92 & 0.346$\pm$0.013 & \(8.76_{-0.18}^{+0.19}\) & 2.95$\pm$0.06 & \(13.23_{-0.03}^{+0.02}\)    \\
\hline
$1.5 \le z < 2.0$ \\
$\rm RS$ & 2066 & 1.731 & 10.41 & 10.69 & 0.195$\pm$0.030 & \(7.87_{-0.70}^{+0.65}\) & 3.10$\pm$0.24 & \(12.94_{-0.13}^{+0.11}\)    \\
$\rm GV$ & 506 & 1.735 & 10.50 & 10.69 & 0.151$\pm$0.037 & \(7.37_{-1.06}^{+0.95}\) & 2.73$\pm$0.33 & \(12.74_{-0.22}^{+0.18}\)    \\
$\rm BC$ & 1171 & 1.782 & 10.53 & 10.69 & 0.146$\pm$0.015 & \(6.31_{-0.37}^{+0.35}\) & 2.59$\pm$0.13 & \(12.60_{-0.10}^{+0.08}\)    \\
\hline
$2.0 \le z \le 2.5^{\rm i}$ \\
%$\rm RS$ & 1251 & 2.231 & 10.00 & 10.54 & 0.294$\pm$0.042 & \(12.21_{-1.00}^{+0.94}\) & 5.50$\pm$0.39 & \(13.36_{-0.10}^{+0.09}\)   \\
$\rm GV$ & 1029 & 2.243 & 10.00 & 10.46 & 0.147$\pm$0.022 & \(7.56_{-0.65}^{+0.61}\) & 3.57$\pm$0.27 & \(12.75_{-0.12}^{+0.11}\)   \\
$\rm BC$ & 2239 & 2.215 & 10.00 & 10.31 & 0.211$\pm$0.025 & \(8.57_{-0.58}^{+0.55}\) & 3.59$\pm$0.21 & \(12.78_{-0.10}^{+0.09}\)   \\
\hline
\end{tabular}}

% Notes: \\
\begin{tabular}{p{18cm}}
$^{\rm a}$ The source number of each galaxy sub-sample. \\
$^{\rm b}$ The median photometric redshift of each galaxy sub-sample. \\
$^{\rm c}$ The stellar mass selection threshold of each galaxy sub-sample. \\
$^{\rm d}$ The median stellar mass of each galaxy sub-sample. \\
$^{\rm e}$ The clustering amplitude of the angular correlation function of each galaxy sub-sample (see Section~\ref{sect:acf}). \\
$^{\rm f}$ The correlation length of the spatial correlation function of each galaxy sub-sample (see Sections~\ref{sect:sc} and \ref{sect:acc_evo}, and Figure~\ref{fig:r0_evo}). \\
$^{\rm g}$ The bias of each galaxy sub-sample (see Section~\ref{sect:bias}). \\
$^{\rm h}$ The halo mass approximately converted from the bias (see Sections~\ref{sect:bias} and \ref{sect:acc_evo}, and Figure~\ref{fig:halomass_evo}). \\
$^{\rm i}$ For a clear comparison, these values are taken from Table~\ref{tab:galaxy_bin} for GV and BC galaxies selected with $\log (M_*/M_\odot) \ge 10$ and at $2.0 \le z \le 2.5$.
\end{tabular}

\label{tab:acc_evo}
\end{table*}

\subsection{The Clustering Similarity between GV galaxies and AGNs}\label{sect:GV_clustering}

\begin{figure}
   \centering
   \includegraphics[width=\columnwidth]{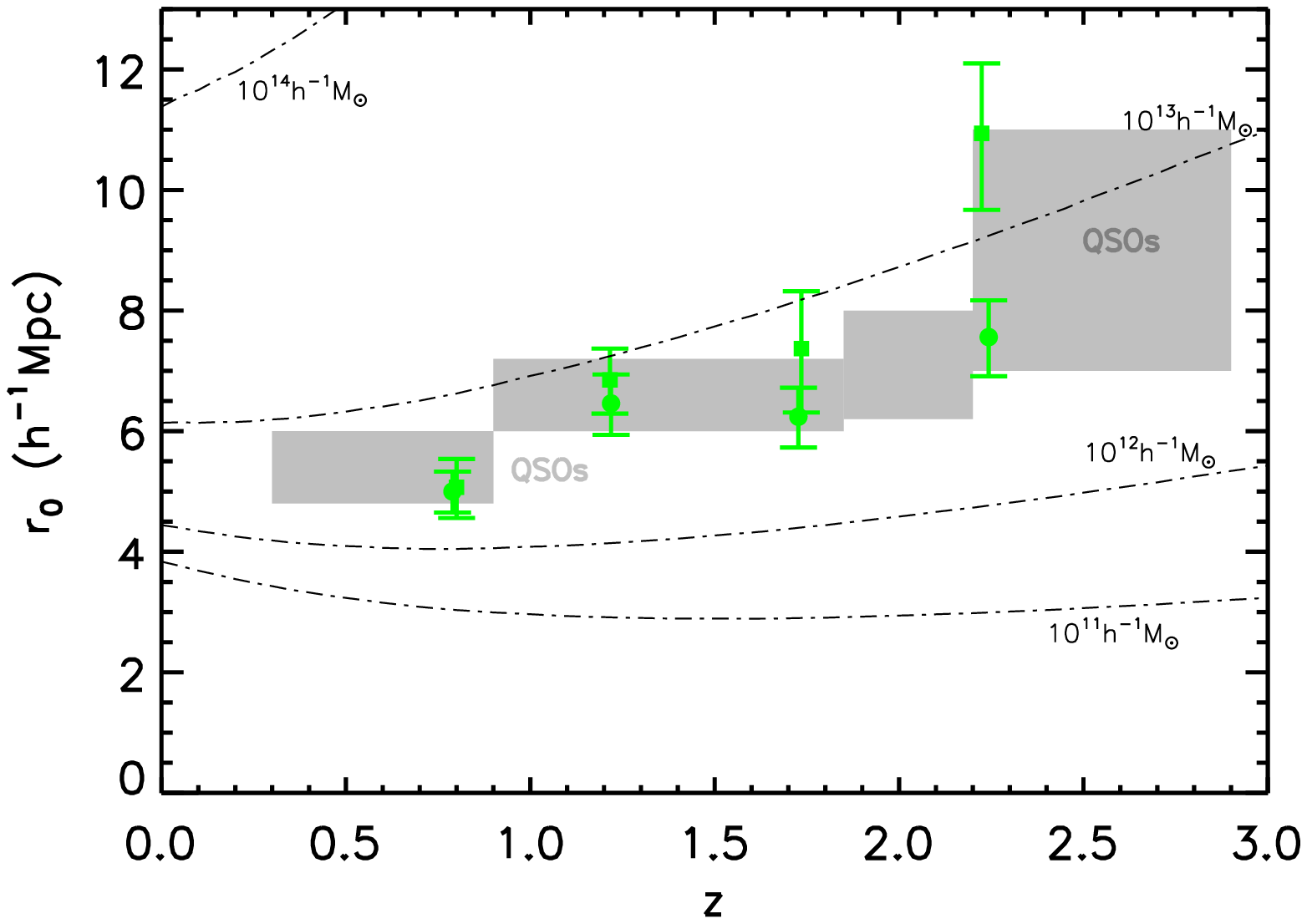}
   \caption{
	Same as Figure~\ref{fig:r0_all}, but only for the massive GV (green) galaxy sub-samples selected with $\log(M_*/M_\odot) \ge 10$ (circles) and 10.5 (squares), compared with the correlation lengths of AGN samples \citep[gray regions;][]{Myers_2006,Ross_2009,Eftekharzadeh_2015}
	}
   \label{fig:r0_GV}
\end{figure}

We note that the correlation lengths of GV galaxies agree well with those of AGN samples in the literature \citep{Myers_2006,Ross_2009,Eftekharzadeh_2015} at all redshifts as shown in Figure~\ref{fig:r0_GV}.

Although the quoted AGN samples from SDSS surveys are mainly selected using magnitude criteria, and then it is different from our mass selection criteria for GV galaxies, the weak dependence of AGN clustering on luminosity has been reported by several works \citep[e.g.,][]{Francke_2007,Shen_2009,Krumpe_2010,Shirasaki_2011,Eftekharzadeh_2015,Ikeda_2015,Mendez_2016}.
For example, \citet{Eftekharzadeh_2015} found that in optical bands quasar clustering over $2.2 \le z \le 2.8$ remains similar over a decade in luminosity, and \citet{Mendez_2016} found a comparable clustering between high- and low-luminosity X-ray/radio-selected AGNs at $0.2 \le z \le 1.2$.
Instead, for our GV galaxies, the clustering weakly depends on the stellar mass completeness at $z \le 2$, while the dependence of GV clustering on the stellar mass potentially emerges at $z \ge 2$, which is in line with the larger range of AGN clustering at $z \ge 2$.
Consequently, we conclude that the clustering of GV galaxies is similar to that of AGNs over $0.5 \le z \le 2.5$.

This similarity is consistent with the results that AGN signatures are truly detected in GV galaxies \citep[e.g.,][]{Pentericci_2013,Law-Smith_2017} and that a higher fraction of AGN is found in GV galaxies than RS and BC galaxies at $z \le 2$ \citep[e.g.,][]{Treister_2009,Hickox_2009,Wang_2017,Gu_2018}.
Therefore, this similarity may suggest that AGN activity and then the role of AGN quenching play an important role in transforming galaxy colors.
The AGN may be triggered by some internal or external processes like gas rich mergers \citep{Springel_2005} or disk instabilities \citep{Elmegreen_2008}.

\section{Summary}\label{sect:summary}

Using the multi-wavelength data in Ultra-Vista COSMOS field, we present a clustering measurement of the UV-selected massive galaxies at $0.5\le z\le2.5$. The large survey area (1.62 ${\rm deg}^2$) and the depth of the survey ($K_{s}=23.4$) enable us to measure the clustering of COSMOS/UltraVISTA galaxies down to stellar mass of $\approx 10^{10}~M_{\odot}$. We use the criteria of \citet{Wang_2017} to define the RS, GV, and BC galaxies up to $z\sim2.5$, before estimating their clustering amplitudes.

The clustering of RS galaxies generally follows a model of a constant halo mass at least from $z\sim2$. This means that originally less massive galaxies grow and join the RS population and that the RS galaxies do not accumulate much stellar mass since then.

The clustering amplitude of GV galaxies generally lies between those of RS and BC galaxies at all redshifts. At $z>1.5$, the clustering of GV galaxies is closer to that of BC galaxies, while at $1.0<z<1.5$, the clustering of GV galaxies is closer to that of RS galaxies. At $z<1$, the clustering of RS galaxies shows an enhancement compared with BC and GV galaxies, which may be due to the rapid quenching of star formation at $z\sim1$ for the massive SFGs.

The GV galaxies and AGNs share similar clustering amplitudes at all redshifts, indicating that AGN activity might be responsible for transforming galaxy colors.

We check the effect of evolution of stellar mass on clustering amplitude, assuming a median halo mass growth rate and a constant SHMR.  We find that the clustering amplitudes of galaxy samples with different colors are all similar once they have a similar median stellar mass. Therefore, we conclude that, for most massive galaxies, the median stellar mass is a good predictor to determine galaxy clustering amplitudes.

\section{acknowledgments}

We are grateful to the referee for her/his many constructive comments that led us to substantially revise our paper.
We acknowledge the constructive work by Adam Muzzin presenting the catalog in the COSMOS-UltraVista field.
This work is supported by the National Natural Science Foundation of China (NSFC, Nos. 11673004, 23002601, 1320101002, 11433005, 11421303, and 11503024), the National Basic Research Program of China (973 Program)(2015CB857004) and the Specialized Research Fund for Shandong Provincial Key Laboratory (KLWH201807).

\bibliography{Reference}
%\bibliography{RGB_clustering.bbl}
%\bibliography{ms.bbl}

\end{document}